\documentclass[aps,pra,twocolumn,superscriptaddress,showpacs,floatfix]{revtex4}
\usepackage{graphicx,dcolumn,bm,hyperref,amsmath,amssymb,xspace,epsfig,float,array,multirow,amsfonts}
\usepackage[vcentermath]{youngtab}
\usepackage{relsize,exscale}
\usepackage{amsmath}
\usepackage{graphicx}
\usepackage{gensymb}
\usepackage[colorinlistoftodos]{todonotes}
\usepackage{color}
\usepackage[english]{babel}
\usepackage[utf8]{inputenc}
\usepackage{tikz,pgfplots}
\usepackage{amsmath}
\usepackage{graphicx}
\usepackage{wasysym}
\usepackage[colorinlistoftodos]{todonotes}
\usepackage{mathtools, bm}
\usepackage{amssymb, bm}
\usepackage{amsmath}

\begin{document}

\title{Bosonic Double Ring Lattice Under Artificial Gauge Fields}

\author{Nicolas Victorin}
\affiliation{Univ. Grenoble Alpes, CNRS, LPMMC, F-38000 Grenoble,  France}

\author{Frank Hekking}\thanks{Deceased on may 15th (2017)}
\affiliation{Univ. Grenoble Alpes, CNRS, LPMMC, F-38000 Grenoble,  France}

\author{Anna Minguzzi}
\affiliation{Univ. Grenoble Alpes, CNRS, LPMMC, F-38000 Grenoble,  France}

\date{\today}

\begin{abstract}
We consider a system of weakly interacting bosons confined on a planar double ring lattice subjected to two artificial gauge fields. We determine its ground state  by solving coupled discrete non-linear Schrödinger equations at mean field level. At varying inter-ring tunnel coupling, flux and interactions  we identify the  vortex, Meissner and biased-ladder phases also predicted for a bosonic linear ladder by a variational Ansatz. We also find peculiar features associated to the ring geometry, in particular parity effects in the number of vortices, and the appearance of a single vortex in the Meissner phase.  We show that the persistent currents on the rings carry  precise information on the various phases. Finally, we propose a way of observing the Meissner and vortex phases via spiral interferogram techniques.
\end{abstract}


\maketitle

\maketitle
\section{Introduction}
Experimental  progress  with  ultracold  quantum  gases  has  made  possible to  engineer the
coupling  between different internal states  of  the  atoms,  and to  realize  synthetic  gauge
fields ~\cite{RevModPhys.83.1523,PhysRevLett.107.255301StrongMagneticFieldsOpticalLattice,PhysRevLett.111.185301Butterfly}. When a neutral atom moves in a properly designed laser field, its center-of-mass motion mimics the dynamics of a charged particle in a magnetic field, under the effect of a Lorentz-like force. The corresponding Aharonov-Bohm phase is related to the Berry’s phase that emerges
when the atom adiabatically follows one of the dressed states of the atom-laser interaction ~\cite{RevModPhys.83.1523}. These advances allow the quantum simulation of a wide range
of Hamiltonians, in particular relevant in condensed matter physics.
Indeed,  some  of  the  most  intriguing  phenomena  in  condensed  matter  physics  involve  the
presence  of  strong magnetic fields. For instance,  topological states  of matter are realized in quantum Hall systems, which are insulating in the bulk, but bear conducting edge states~\cite{QuantumHall}. 

A ladder is  the simplest geometry where one can get some insight on two-dimensional quantum system subjected to a synthetic gauge field ~\cite{PhysRevX.7.021033,PhysRevB.92.115446}.  The bosonic linear ladder has been the subject of intense theoretical work. The phase diagram has been established by means of field-theoretical methods~\cite{Georges,PhysRevB.64.144515Orignac}, and intensive DMRG simulations~\cite{PhysRevB.91.140406DMRG}. Those studies, in addition to common features of Bose-Hubbard models such as superfluid and Mott insulating phases, revealed new exciting phase of matter induced by the magnetic field: chiral superfluid phases, chiral Mott insulating phases displaying Meissner currents ~\cite{PhysRevB.64.144515Orignac,ChiralMottDhar} and vortex-Mott insulating phases~\cite{PhysRevLett.111.150601}. In the weakly interacting regime, on which we will focus on this work, an additional phase has been predicted ~\cite{Mueller} a biased ladder phase characterized by an imbalanced population of the bosons between the two legs, explicitly breaking $\mathbb{Z}_2$ symmetry. This phase was shown to be stable in the  interacting case, except for a special value of the applied flux, where   umklapp processes destabilize it ~\cite{PhysRevA.92.013625Uchino}. The dependence of the critical flux separating Meissner and vortex phase on interparticle interactions has been also studied ~\cite{Oktel}. In parallel to these theoretical advances, the experimental realization of the bosonic flux ladder has been reported in optical lattices~\cite{Atala} as well as for lattices in  synthetic dimensions, both for fermions and bosons ~\cite{Stuhl,Mancini}.

In this work, we consider a system made of two one-dimensional coupled lattice rings subjected to different flux in each leg.   This specific bosonic ladder corresponds to different boundary conditions than the case of a linear ladder.  In particular, this double ring lattice geometry allows to study persistent currents in dimension larger than one~\cite{Marco}, which shows promising applications for atomtronics developments~\cite{Atomtronics,LuigiD}.  At difference from~\cite{Qubit,Tobias}, we focus on a planar geometry with concentric rings, as could be realized eg with dressed potentials~\cite{Helene}, or  using copropagating Laguerre-Gauss beams~\cite{CylindricalOpticalLattice}. 

We study first the properties of the non interacting gas. After identifying the vortex and Meissner phases, we discuss specific features of the double ring lattice  geometry, as the  appearance of a vortex in the Meissner phase and parity effect in the vortex phase,  and the behavior of persistent currents.
 
Through a numerical study we then explore the dilute, weak-interacting regime and address the nature of the ground state at mean field level. In particular we identify known  phases such as the Meissner, vortex  and biased-ladder phases \cite{Mueller} as well as the effect of commensurability of the total flux.
Finally, we propose the spiral interferogram images -- obtained by interference among the two rings during  time of flight expansion --  as a probe of vortex-carrying phases, specifically adapted to the ring geometry.

\section{The model}
\begin{figure}[hbtp]
\centering
\includegraphics[scale=0.4]{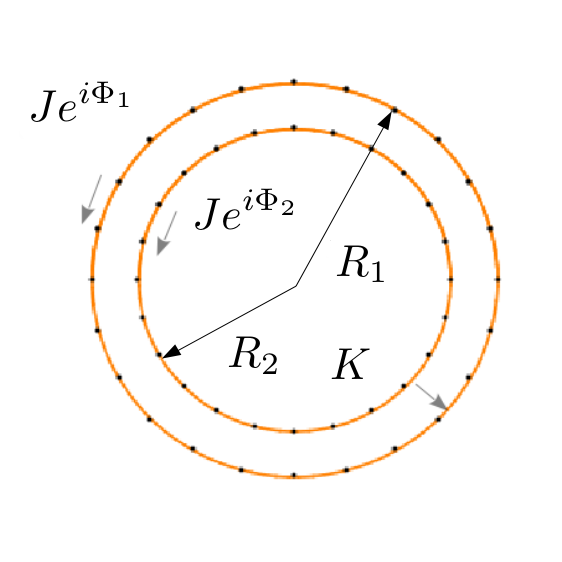}
\caption{(Color online) Representation of the geometry studied in this work: coplanar ring lattices of radii $R_1$ and $R_2$ with the same number of sites, with inter-ring tunnel energy $K$ and intra-ring tunnel energies $J e^{i \Phi_p}$, with $p=1,2$.}
\label{fig1b}
\end{figure}
We consider a Bose gas confined in a double ring lattice. In the tight-binding approximation we model the system using the Bose-Hubbard model:
\begin{align}
&\hat{H}=\hat{H}_0+\hat{H}_{int}=\nonumber\\ &-\!\sum_{l=1,p=1,2}^{N_s} \!\!\!J_p\left(a^{\dagger}_{l,p}a_{l+1,p}e^{i\Phi_p} + a^{\dagger}_{l+1,p}a_{l,p}e^{-i\Phi_p}\right)\nonumber \\ &-\!K\sum_{l=1}^{N_s}\left(a_{l,1}^{\dagger}a_{l,2}+a_{l,2}^{\dagger}a_{l,1}\right)+\frac{U}{2}\!\!\!\sum_{l=1,p=1,2}^{N_s}a^{\dagger}_{l,p}a^{\dagger}_{l,p}a_{l,p}a_{l,p}
\label{eq1}
\end{align}
where the angular position on the double ring lattice is given by $\theta_l=\frac{2\pi}{N_s}l$ where $l$ is an integer $l \in \left[1, N_s\right]$ with $N_s$ the number of sites in each ring. 
In Eq (\ref{eq1}) $J_1$ and $J_2$ are respectively the tunneling amplitude from one site to an other along each ring, the parameter $K$ is the tunneling amplitude between the two rings, connecting only sites with the same position index $l$ and $\Phi_{1,2}$ are the fluxes threading the inner and outer ring respectively. In the case where the gauge fields are induced by applying a rotation to each ring one has $\Phi_i=\frac{2\pi}{N_s}\frac{\tilde{\Phi}_i}{\Phi_0}$, with $\tilde{\Phi}_i=\Omega R_i^2$, $\Omega$ being the angular rotation frequency, $R_i$ radius of ring $i$, $\Phi_0=2\pi \hbar/m$ the Coriolis flux quantum. As $J_i \approx \frac{\hbar^2}{2mR_i^2}$,  to lowest order we can consider $J_1 \approx J_2$ corresponding to two rings close to each other, or realized using adjusted lattice potential.  In the following, it will be useful to introduce the relative flux $\phi = \Phi_1-\Phi_2$ and average flux $\Phi = (\Phi_1+\Phi_2)/2$.

\section{Non interacting regime}
We first proceed by analyzing the non-interacting problem.
The diagonalization of $H_0$ (see Appendix A for details) yields the following two-band Hamiltonian:
\begin{eqnarray}
\hat{H}_0 = \sum_k \alpha_k^{\dagger}\alpha_kE_+(k) + \beta_k^{\dagger}\beta_k E_-(k),
\end{eqnarray}
where
\begin{align}
\begin{pmatrix}
a_{k,1}\\
a_{k,2}
\end{pmatrix}
=
\begin{pmatrix}
v_k & u_k\\
-u_k & v_k
\end{pmatrix}
\begin{pmatrix}
\alpha_k\\
\beta_k
\end{pmatrix},
\end{align}
and the functions $u_k$ and $v_k$ depend on the parameters $\phi$ and $K/J$ (see Appendix A for details), the momentum in units of inverse lattice spacing takes discrete values given by  $k =\frac{2 \pi n}{N_s}$, with $n=0,1,2... N_s-1$  and the dispersion relation    $E_{\pm}(k)$ reads 
\begin{align}
E_{\pm}(k)=-&2J\cos(\phi/2)\cos(k-\Phi)\nonumber\\ \pm &\sqrt{K^2+(2J)^2\sin(\phi/2)^2\sin(k-\Phi)^2}.
\label{eq:epm}
\end{align}
We see that the only influence of the average flux $\Phi$ is to shift in momentum space the energy spectrum.

\begin{figure}[h!]
\includegraphics[scale=1]{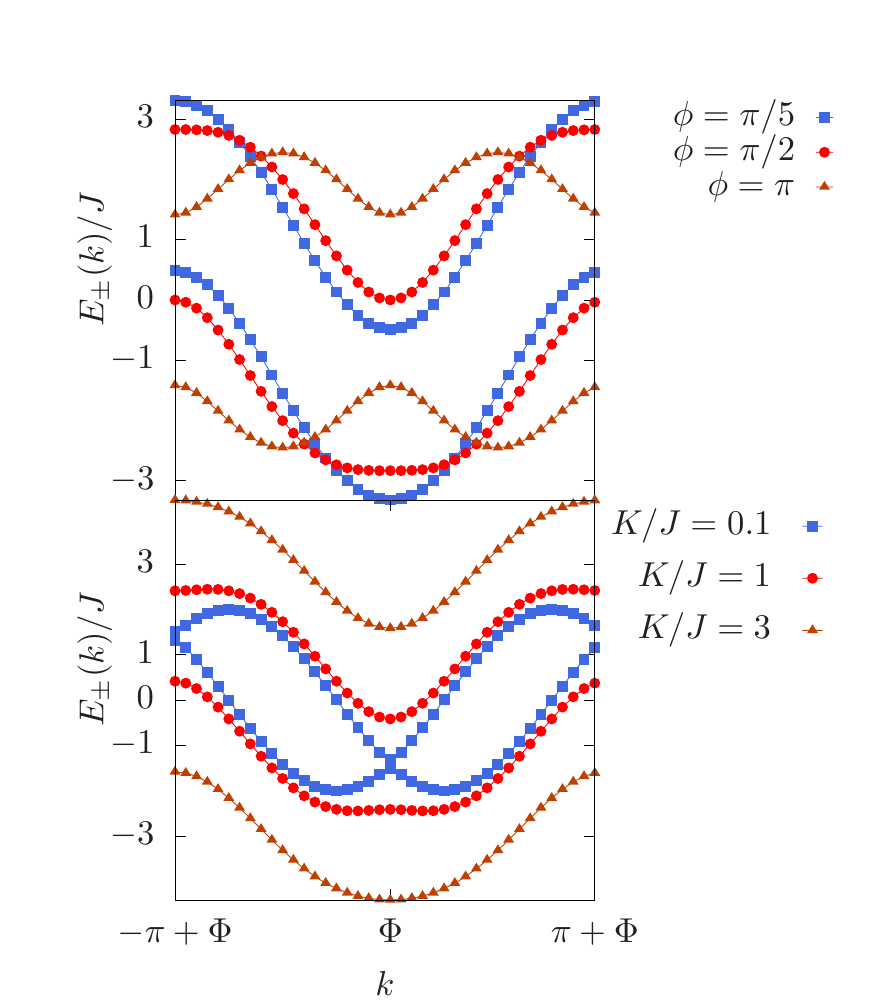}
   \caption{(Color online) Energy spectrum (in units of $J$ with $N_s=40$ sites on each ring) as a function of wavevector $k$ (in units of inverse lattice spacing)  of non-interacting bosons on a double ring lattice, for several values of the tunneling ratio $K/J$ at fixed relative flux $\phi=\pi/2$ (bottom) and several values of $\phi$ at fixed $K/J=\sqrt{2}$ (top).}
      \label{fig1}
   \end{figure}

\begin{figure}
 \begin{tikzpicture}
    \matrix [ampersand replacement=\amp] {
     \node{\includegraphics[scale=0.2]{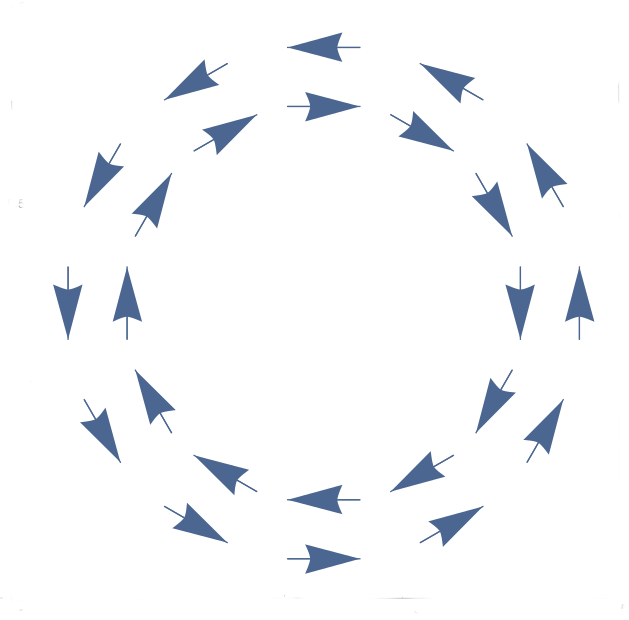}}; \amp \node[text height=1.5ex,text depth=.25ex]{$\Phi=0$, $K/J=2$}; \\
      \node{\includegraphics[scale=0.3]{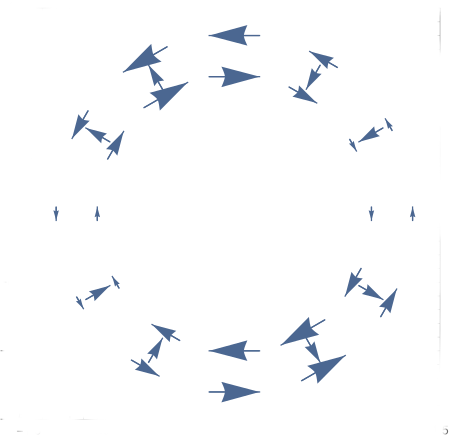}}; \amp \node[text height=1.5ex,text depth=.25ex]{$\Phi=0$, $K/J=0.9$ }; \\
      \node{\includegraphics[scale=0.2]{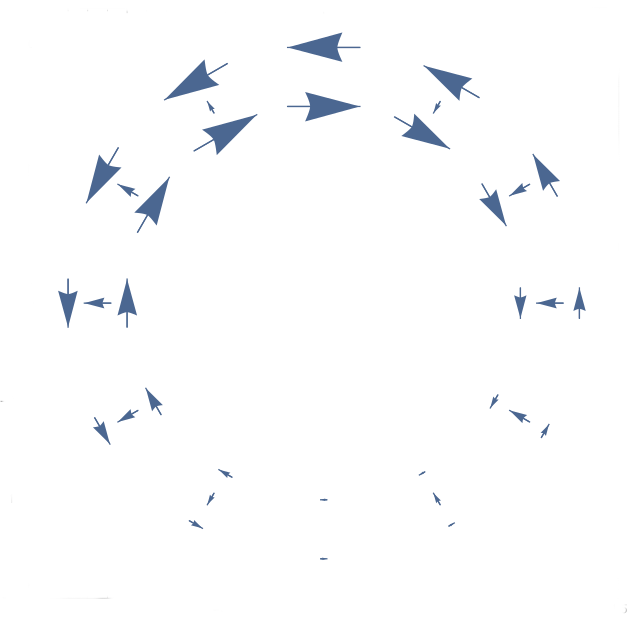}}; \amp \node[text height=1.5ex,text depth=.25ex]{$\Phi=\frac{\pi}{N_s}$,$K/J=0.9$}; \\
    };
  \end{tikzpicture}
 \caption{(Color online) Representation of the current patterns for noninteracting bosons on a double ring lattice in various parameter regimes as indicated on the figure.  The length of arrows is proportional to the amplitude of the current field.  The currents fields are minimal at the core of the vortex, where also the density drops. Upper panel: Meissner phase. Middle panel: vortex phase, case of two vortices. Lower panel: single vortex in the Meissner phase.  In all panels, $\phi=\pi/2$ and   $N_s=12$.}
\label{figcurrents}
\end{figure}
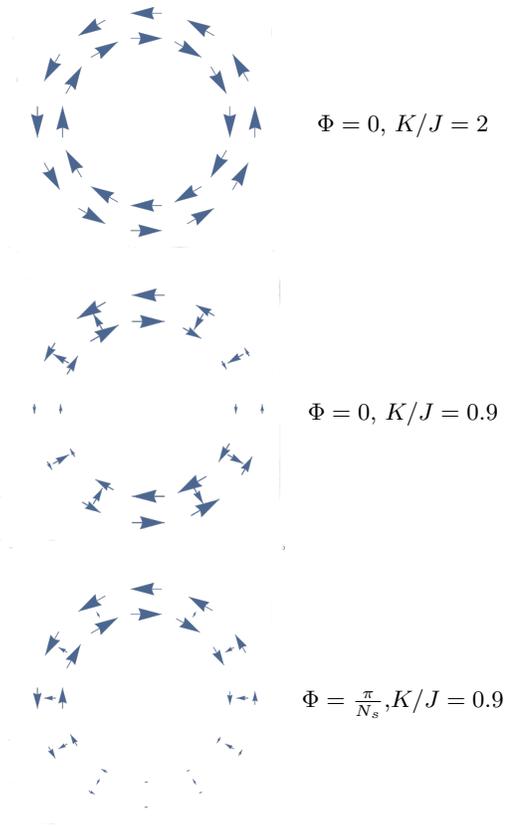

The relevant ground state properties are obtained from the low-energy branch spectrum since, for a finite size-ring,  at $T=0$ and $U=0$ the bosons form a condensate in the lowest-energy state available. At varying tunneling ratio $K/J$ and relative flux $\phi$, two possible situations arise from the lowest-energy branch $E_-(k)$ (see Fig.~\ref{fig1}). When $E_-(k)$ has a single minimum, the bosons condense in the state $k=\Phi$, corresponding to the Meissner phase, while one has a vortex phase when $E_-(k)$ has two minima and bosons condense with the same occupancy in each of the two minima $k_1$ and $k_2$ given by
\begin{align}
k_{1,2}= \Phi \mp \arccos\left[\cot\left(\frac{\phi}{2}\right)\sqrt{\left(\frac{K}{2J}\right)^2+\sin^2\left(\frac{\phi}{2}\right)}\right].
\label{eq:k}
\end{align}
	Other possible occupancies of the two minima are discussed in Appendix A.
   The vortex to Meissner phase transition has been experimentally observed in bosonic linear flux ladders ~\cite{Atala}. At fixed $K/J$ value,   the critical flux where the transition appears is obtained by determining the change of curvature in $E_-(k=\Phi)$, thus yielding ~\cite{Georges}:
    \begin{align}
    \phi_c = 2\arccos\left[\sqrt{\left(\frac{K}{4J}\right)^2+1}-\left(\frac{K}{4J}\right)\right].
    \end{align}
   The Meissner phase is characterized by vanishing transverse currents $j_{l,\perp}= iK\langle a^{\dagger}_{l,1}a_{l,2}-a^{\dagger}_{l,2}a_{l,1}\rangle$; the longitudinal currents on each ring, defined as $j_{l,p}= iJ\langle a_{l,p}^{\dagger}a_{l+1,p}e^{i\Phi_p}-a_{l+1,p}^{\dagger}a_{l,p}e^{-i\Phi_p}\rangle$, are opposite and the chiral current, i.e $J_c=\sum_l \langle j_{l,1}-j_{l,2}\rangle$ is saturated. The vortex phase is characterized by a modulated density, jumps of the phase of the wave function, and 
non-zero, oscillating  transverse currents which create a  vortex pattern. This is illustrated in Fig.\ref{figcurrents}, which shows the longitudinal and transverse current configurations both in the Meissner and in the vortex phase.

\begin{figure}[h!]
\includegraphics[scale=0.8]{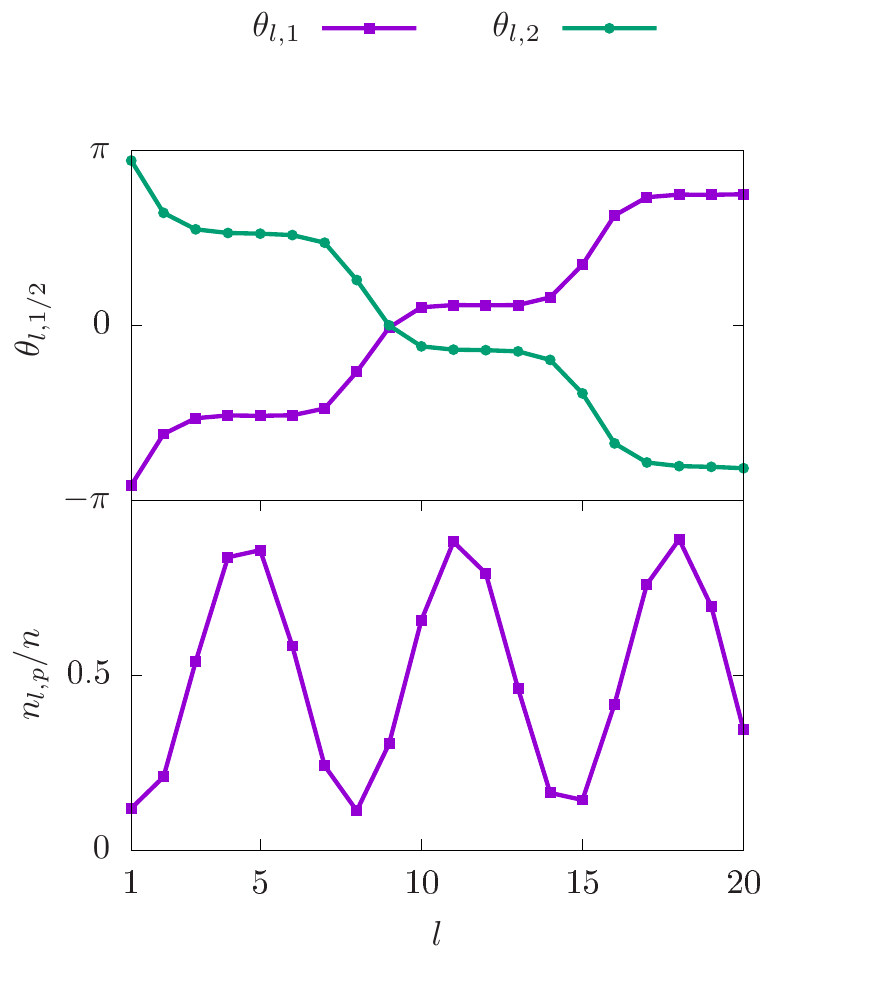}
\includegraphics[scale=0.8]{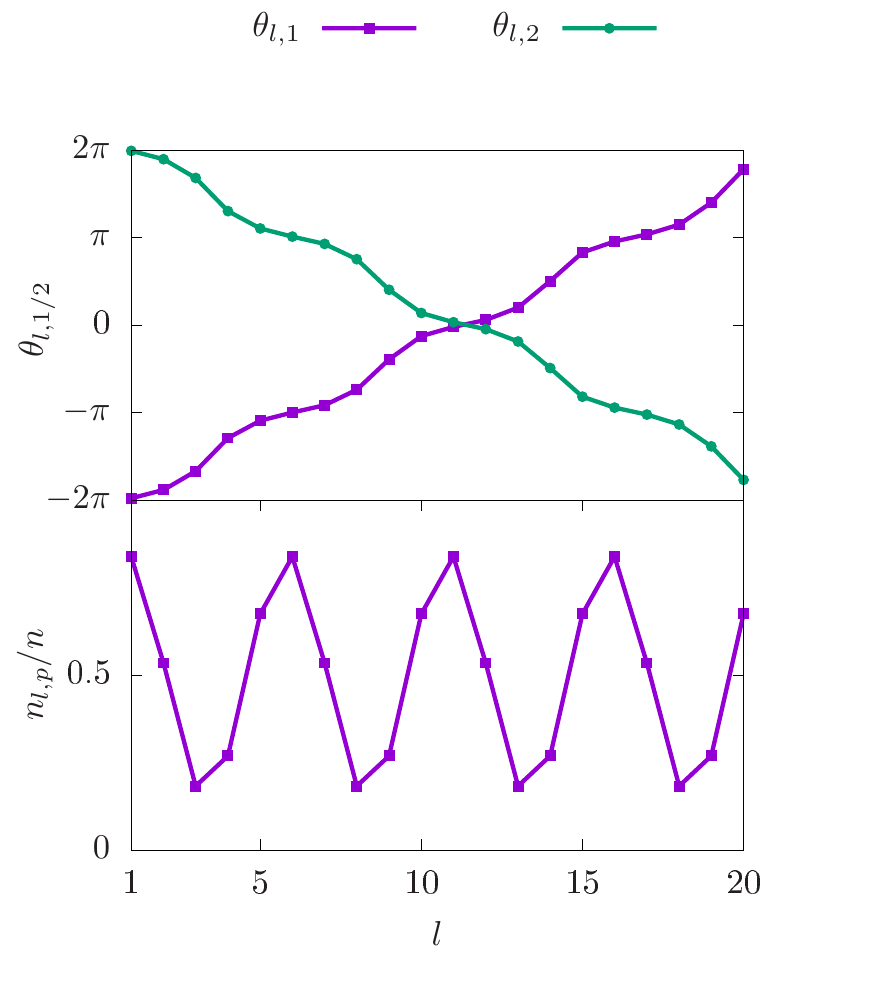}
\caption{(Color online) Phase and density profiles of the condensate wavefunction for noninteracting bosons along the double ring lattice as a function of the lattice index. Top panel:  odd number of vortices for  average flux $\Phi=\pi/N_s$. Bottom panel:  even number of vortices for $\Phi=0$. The other parameters are $K/J=0.8$, $\phi=\pi/2$,$N_s=20$ and $n=N/N_s$.}
\label{fig:vortices}
\end{figure}

\subsection{Vortex configurations on a finite double ring lattice}
\label{vortex-subsec}

\begin{figure}[h!]
\includegraphics[scale=1]{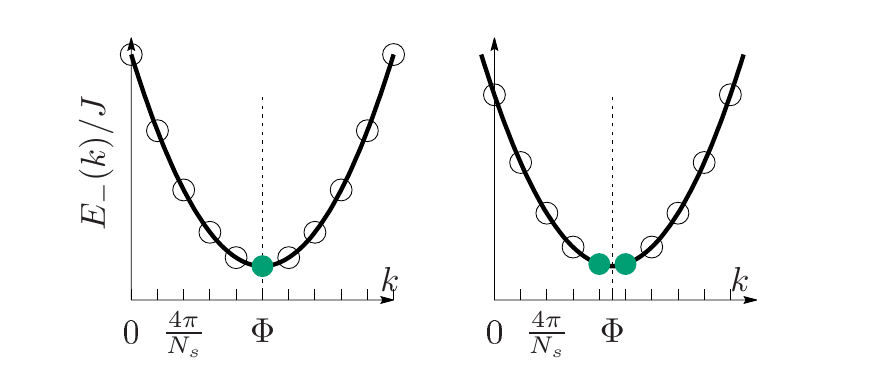}
\includegraphics[scale=1]{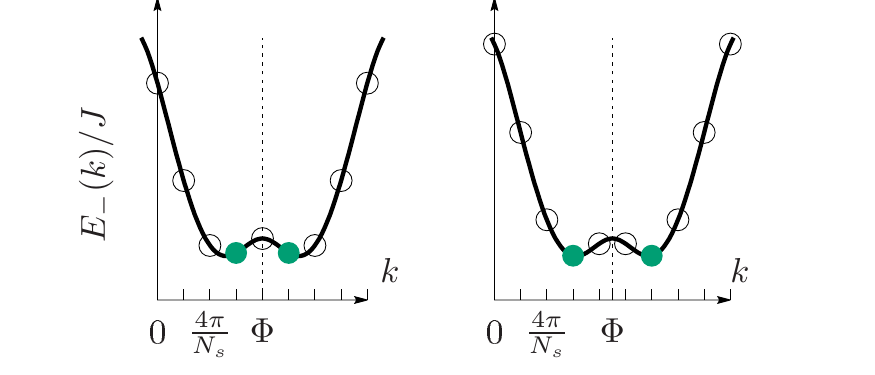}
\caption{(Color online) Scheme of the occupancy of the single-particle levels by noninteracting bosons at zero temperature (filled green circles), on the single-particle dispersion relation in the energy-momentum plane (empty circles joined by line), for various choices of total flux $\Phi$ (dashed vertical line).  In the Meissner phase, when $\Phi=2j \frac{\pi}{N_s}$ (left top panel, with $\Phi=10\pi/N_s$) bosons condense in the $k=\Phi$ mode. When  $\Phi=(2j+1)\frac{\pi}{N_s}$  (right top panel, with $\Phi=9\pi/N_s$), $\Phi$  lies between two momentum modes, the lowest-energy states are doubly degenerate and the system supports a vortex in the Meissner phase. In the vortex phase, when  $\Phi=2j\frac{\pi}{N_s}$  (bottom left panel, with $\Phi=8\pi/N_s$) we find an even number of vortices, whereas when $\Phi=(2j+1)\frac{\pi}{N_s}$  (bottom right panel, with $\Phi=9\pi/N_s$) the number of vortices is odd. Notice that the scheme is completely general for values of  $\Phi$ equal to any odd or even multiple of $\pi/N_s$.}
\label{fig3}
\end{figure}

Figure~\ref{fig:vortices} shows the distribution of the phase and density of the condensate wave function of the noninteracting gas in the vortex phase, which reads $\psi_{l,p}=\sqrt{\frac{N}{2N_s}}\left(\delta_{p,1}(u_{k_1}e^{ik_1l}+u_{k_2}e^{ik_2l})+\delta_{p,2}(v_{k_1}e^{ik_1l}+v_{k_2}e^{ik_2l})\right)$, for various values of the system parameters. The number  $N_v$ of vortices is obtained by counting the number of jumps in the phase. Since it is also associated to the number of oscillations in the density, which are characterized by the wavevector  $k=k_2-k_1$, it is readily obtained as $N_v= N_s (k_2-k_1)/2\pi$.  Recalling that  the value of the total flux $\Phi$ fixes the position of the minima of the dispersion relation (\ref{eq:epm}), in the case where  the total flux is multiple of $\frac{\pi}{N_s}$ we obtain specific features associated to the commensurability of $\Phi$ with the allowed values of the discrete wavevector $k$.  Figure~\ref{fig3} depicts the various possibilities. When $\Phi=2 j\frac{\pi}{N_s}$, with $j$ integer number,   the dispersion relation is centered on an allowed  value of the  quantized momentum $k$. In this case vortices start to form when the dispersion relation displays a double-minima structure, and  the number of vortices is even.  

On the other hand, when 
 $\Phi = (2j+1)\frac{\pi}{N_s}$  the value of $\Phi$ falls among two adjacent values of quantized momentum $k$ (see again Fig.\ref{fig3}). In this case, in the vortex phase, the distance among the two minima corresponds to an odd multiple of $\frac{\pi}{N_s}$, giving rise to an odd number of vortices.  Quite interestingly, in the Meissner phase, ie for a choice of parameters $\phi$ and $K/J$  leading to  a single minimum in the single-particle excitation dispersion $E_-(k)$,  for  $\Phi = (2j+1)\frac{\pi}{N_s}$  we find a nontrivial  pattern in the current profiles, corresponding to a single vortex configuration (see Fig.\ref{figcurrents}, third panel). This is a mesoscopic effect associated to the finite size and the geometry of the ring. As we shall see below, however, this vortex is more fragile than those appearing in the vortex phase, and is destroyed in the presence of interactions.

\subsection{Persistent and chiral currents}
\begin{figure}[h!]
\includegraphics[scale=1]{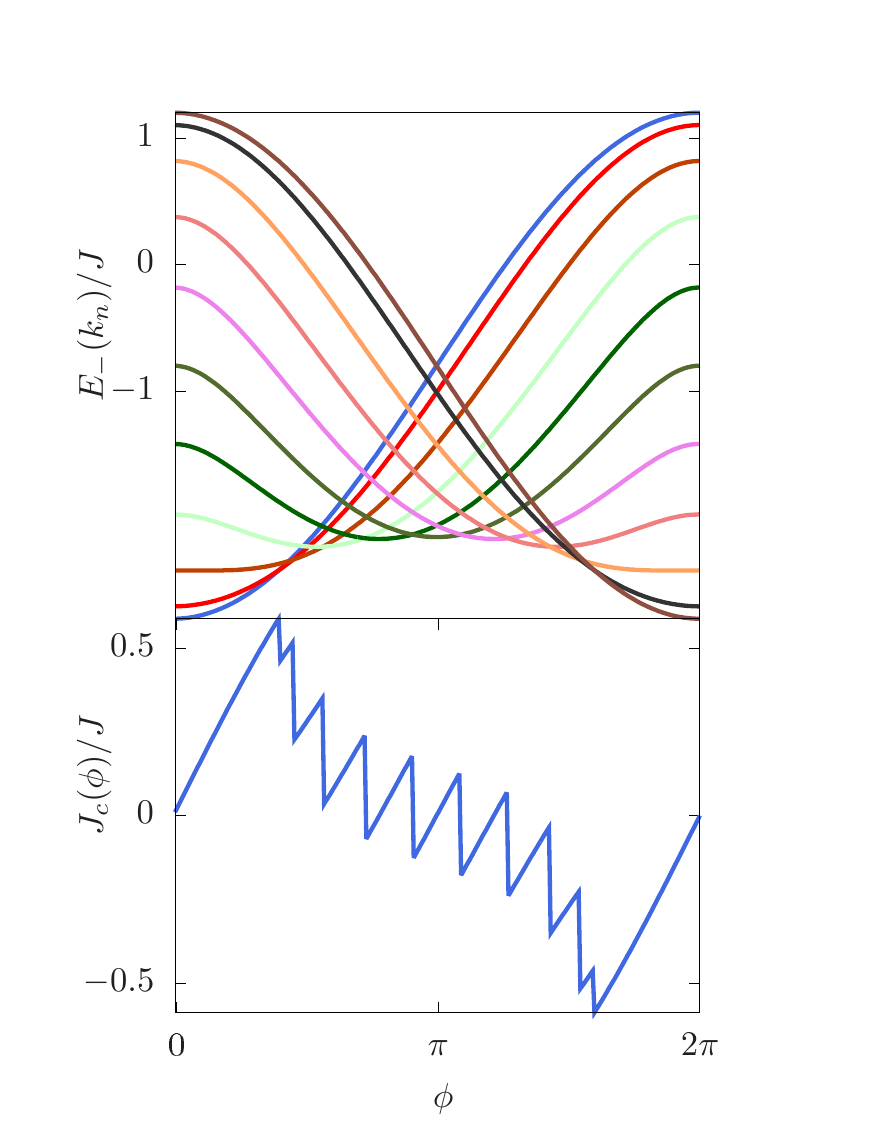}
\caption{(Color online) Upper panel: Excitation branches  $E_-(k_n,\phi)$ as a function  of the relative flux $\phi$ (dimensionless)  for various values of $k_n=\frac{2\pi}{N_s}n$, $n\in [0,N_s/2]$. At $\phi=0$, one has  $E_-(k_0, \phi)<E_-(k_1,\phi)<\dots<E_-(k_n, \phi)$ (blue to brown curves, from bottom to top). The energy of the lowest excitation branch is the lower envelope of these curves and is used to calculate the chiral current.   Lower panel: chiral current, obtained from Eq.~(\ref{jchiral}), as a function of  $\phi$.
In both panels we have taken  $N_s=20$, $\Phi=0$ and $K/J=0.8$.
}
\label{fig4}
\end{figure}

We proceed next to study the persistent currents on the ring. They are defined as $I_p=\frac{\partial\langle H\rangle}{\partial \Phi_p}$. Since for the Hamiltonian (\ref{eq1}) one has  $\frac{\partial\langle H\rangle}{\partial \Phi}=0$, we obtain that $I_1=-I_2=I$ and we have a correspondence between chiral  current $J_c$ and persistent current:
\begin{equation}
J_c=2I=\frac{\partial \langle H\rangle}{\partial \phi}.
\label{jchiral}
\end{equation}
 In particular, Fig~\ref{fig4} represents the dependence of the excitation spectrum branches on the relative flux. In order to obtain the persistent currents for each value of $\phi$ we identify the lowest-energy branch as defined piece-wise by following the lowest-energy part of $E_-(k)$ (see Fig.~\ref{fig4} upper panel). The persistent current is then readily obtained by deriving this curve with respect to the flux $\phi$.
The resulting persistent current as a function of relative flux $\phi$ is illustrated in Fig.~\ref{fig4} (bottom panel). By increasing the relative flux at fixed $K/J$, the system undergoes a transition from Meissner to vortex phase. For low $\phi$ values, the particle stays in the branch $E_-(k=\Phi)$ as long as it is in the Meissner phase. At the critical value $\phi_c$ for entering the vortex phase, the persistent current displays a jump, and takes an angular momentum value equal to $\Phi+2\pi/N_s$. As the flux $\phi$ increases, the persistent currents display several other jumps, each corresponding to the appearance of a vortex pair in the ring. We notice that the total number of jumps in the current curve corresponds to $N_s/2$, ie the maximal number of vortex pairs on the ring.

\section{Weakly-interacting regime}

\subsection{Variational Ansatz}
In the case of non interacting bosons, when the single-particle spectrum has two degenerate minima, the many-body ground state energy is highly degenerate as it corresponds to all possible partitions of the particles among the two minima. In the presence of interactions this degeneracy is broken. Introducing the variational Ansatz
\begin{align}
|\Phi_N\rangle=\frac{1}{\sqrt{N!}}\left(\cos(\theta/2)\beta^{\dagger}_{k_1}+\sin(\theta/2)\beta^{\dagger}_{k_2}\right)^N|0\rangle,
\label{Ansatz}
\end{align}
which is valid in weakly interacting regime, Wei and Mueller \cite{Mueller} have identified two phases, corresponding to two different partitions of the bosons on the minima $k_1$ and $k_2$: a vortex phase, when each minimum is occupied by $N/2$ bosons, occuring if $1-6u_{k_1}v_{k_1}>0$; and a biased ladder phase, characterized by symmetry breaking and full occupancy of only one of the two minima, occurring  when $1-6u_{k_1}v_{k_1}<0$. The biased ladder phase is characterized by the absence of  density modulations and different density values on the two rings.

\subsection{Coupled discrete nonlinear Schr{\"o}dinger equations (DNLSE)}

In order to explore in a broader way the weakly-interacting regime, we study the ground state of the system in the mean-field approximation, obtained by  neglecting the quantum fluctuations and correlations. 

We start from the equations of motion for the bosonic field operators in the Heisenberg picture:
\begin{eqnarray}
i\hbar\frac{d a_{l,p}(t)}{dt}=\left[a_{l,p}(t),H\right].
\end{eqnarray}
Taking the mean-field approximation, ie setting $\Psi_{l,p}(t)= \langle a_{l,p}(t)\rangle$
 we obtain two coupled discrete non-linear Schr{\"o}dinger equations (DNLSE): 
\begin{eqnarray}
\label{dnlse}
i\partial_t \Psi_{l,1}(t)&=& -J\Psi_{l+1,1}(t)e^{i(\Phi+\phi/2)}-J\Psi_{l-1,1}(t)e^{-i(\Phi+\phi/2)}\nonumber\\&-&K\Psi_{l,2}(t)+U|\Psi_{l,1}(t)|^2\Psi_{l,1}(t)\\
i\partial_t \Psi_{l,2}(t)&=& -J\Psi_{l+1,2}(t)e^{i(\Phi-\phi/2)}-J\Psi_{l-1,2}(t)e^{-i(\Phi-\phi/2)}\nonumber\\&-&K\Psi_{l,1}(t)+U|\Psi_{l,2}(t)|^2\Psi_{l,2}(t)
\end{eqnarray}
This is the lattice version of the Gross-Pitaevskii equations. The above equations are  expected to hold  for weak interactions and large number of particle on each site.

The corresponding  energy functional is given by
\begin{align}
&E[\mathbf{\Psi}_1,\mathbf{\Psi}_2]=-J\sum_{l,p}\left(\Psi^*_{l,p}\Psi_{l+1,p}e^{i\Phi_p}+c.c\right)\nonumber\\&-K\sum_l\left(\Psi_{l,1}^*\Psi_{l,2}+c.c\right)+\frac{U}{2}\sum_{l,p}|\Psi_{l,p}|^4,
\label{NRJF}
\end{align}
where $\mathbf{\Psi}_p=\{\Psi_{l,p}\}$.

We use a split-step Fourier transform method~\cite{Split} to solve the discrete time dependent NLSE and perform imaginary-time evolution to obtain the ground state of the system with the normalization condition,
\begin{align}
\sum_{l=1}^{N_s}\sum_{p=1,2} |\Psi_{l,p}|^2=N
\end{align}
where $N$ is the number of particles in the system.

\section{Numerical results}

\subsection{Mean field ground state phase diagram}
\begin{figure}[h!]
\includegraphics[scale=0.85]{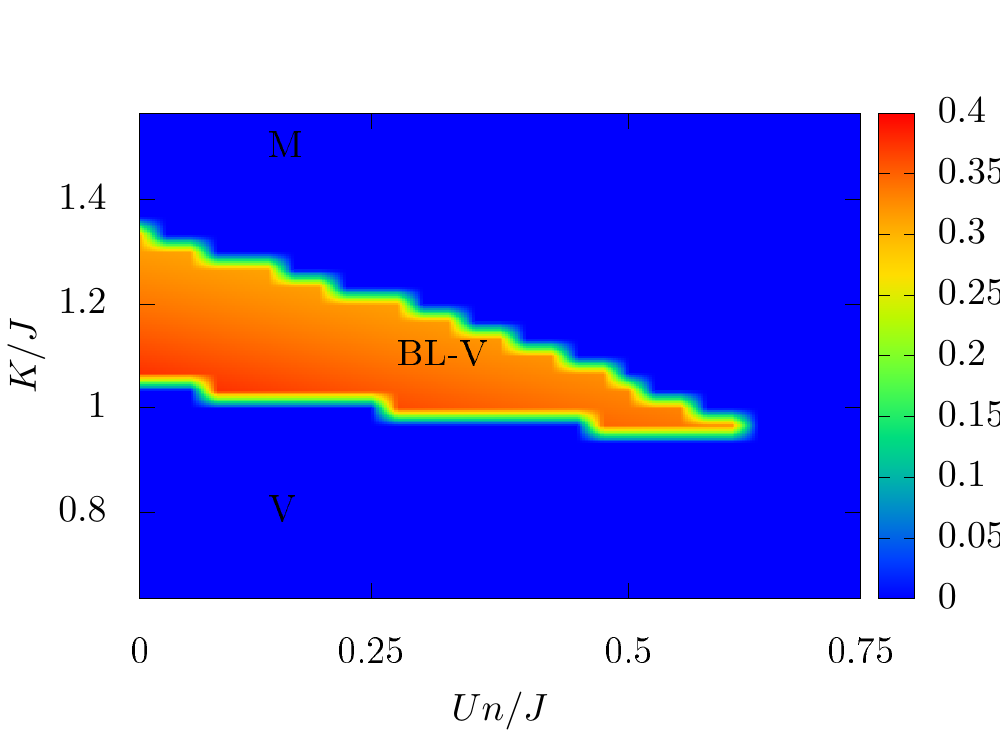}
\includegraphics[scale=0.85]{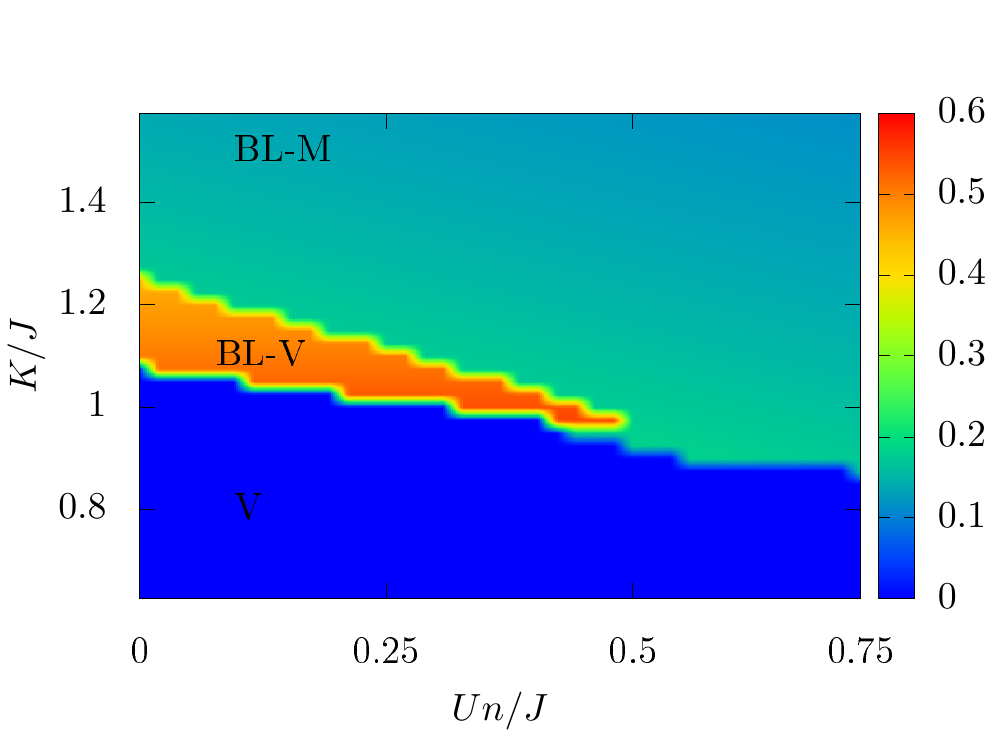}
\caption{(Color online) Color map of the imbalance among particle numbers in each ring,  in the ($K/J$,$Un/J$) plane, for (upper panel) $\phi=\pi/2$, $\Phi=6\pi/N_s$ and $N_s=20$, (lower panel) $\phi=\pi/2$, $\Phi=\pi/N_s$ and $N_s=20$ The letters indicate the parameter regimes where we find a biased-ladder phase  (BL-V) where the single-particle spectrum has a double minimum, a Meissner phase (M), a vortex phase (V) and a  biased-ladder phase (BL-M) where the single-particle spectrum has a single minimum. The  corresponding density profiles are illustrated in Fig.~\ref{Density} and ~\ref{fig15}.}
\label{Nldif}
\end{figure}

\begin{figure}[h!]
\centering
\includegraphics[scale=1]{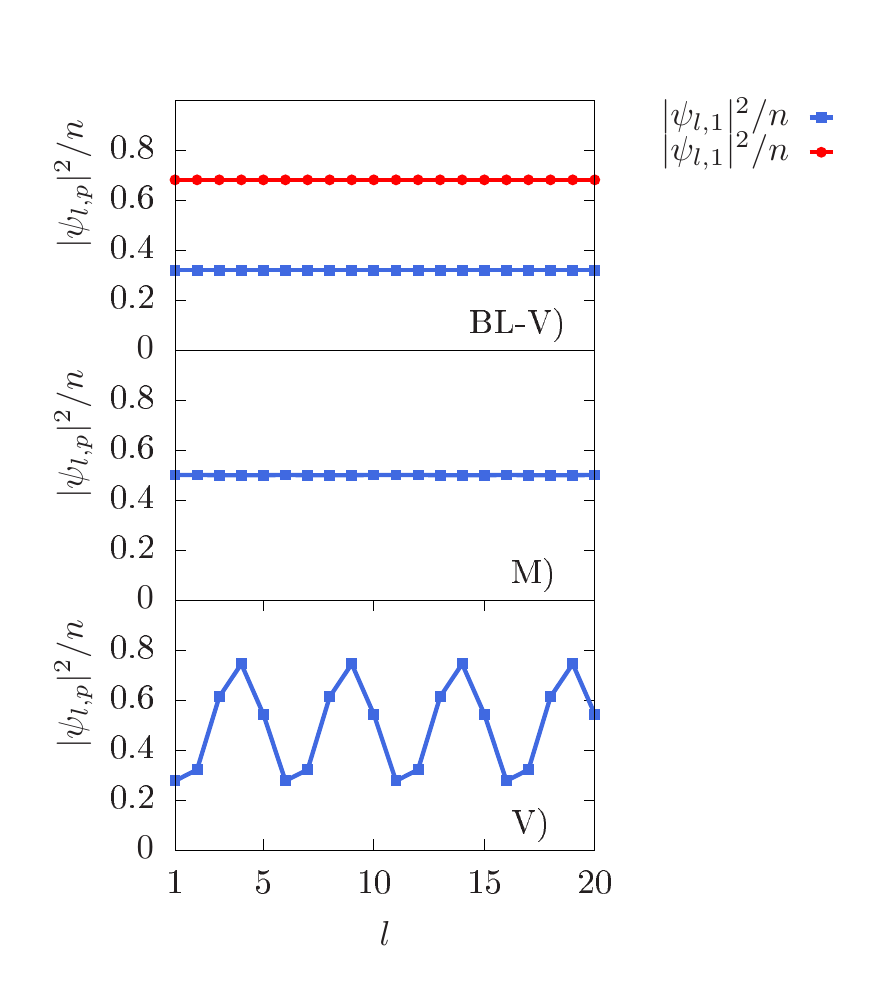}
\caption{(Color online) Density profiles along each ring as a function of the lattice index along the ring, with $N_s=20$, $\phi=\pi/2$, $\Phi=6\pi/N_s$, in the various phases identified in the diagram of Fig.~\ref{Nldif}:
with parameter $\phi=\pi/2$,
 $N_s=20$, $Un/J=0.05$ 
 and biased-ladder phase (BL-V), 
for $K/J=1.1$; Meissner phase (M), for $K/J=2$; vortex phase (V), for $K/J=0.5$ and $Un/J=0.3$.}
\label{Density}
\end{figure}

We use the numerical solution of the DNLSE (\ref{dnlse}) to explore the nature of the ground state at varying interactions and inter-ring tunnel coupling, as identified by the ratios $U n/J$ and $K/J$, with $n=N/N_s$. For simplicity of the analysis, we choose a fixed value  $\phi=\pi/2$ for the relative flux.
Our results are illustrated in Fig.~\ref{Nldif}, showing the particle imbalance among the two rings $\Delta=\left|\sum_l\langle n_{l,1}-n_{l,2}\rangle\right|/N$. For a choice of total flux $\Phi$ corresponding to an even multiple of $\pi/N_s$ (upper panel of Fig.~\ref{Nldif}) at varying interaction and tunnel parameters we identify three phases: the vortex (V) and Meissner (M) phases found in the non-interacting regime, as well as the biased-ladder phase (BL-V) predicted by the variational Ansatz. We have denoted this latter phase  BL-V since it is competing with the vortex phase, and are both obtained from the Ansatz when the single-particle spectrum has a double minimum structure.
Figure ~\ref{Density} shows the corresponding density profiles of the various phases: biased-ladder, Meissner and vortex phases are illustrated in panels (BL-V), (M) and (V) respectively.

For values of total flux corresponding to an odd multiple of $\pi/N_s$ (lower panel of  Fig.~\ref{Nldif}) in place of the Meissner phase admitting a single vortex, as predicted in absence of interactions, we find a biased-ladder phase (denoted as BL-M in the figure).  As it will be  discussed in section V.B, this is a mesoscopic effect due to the finite size of the ring --  the  imbalance decreases with increasing number of sites on the ring.

\subsection{Fate of the single vortex in the Meissner phase}
As discussed in Sec.\ref{vortex-subsec}, in the case when the total flux  $\Phi = (2j+1)\frac{\pi}{N_s}$ and the system is in the Meissner phase, the noninteracting solution predicts  the formation of a single vortex.
We explore here the fate of such a vortex in the presence of weak interactions. 

A first answer is provided by the variational Ansatz introduced in Ref.\cite{Mueller} specialized to the case where the bosons occupy  two neighbouring momentum states of the single-particle excitation spectrum centered around $k=\Phi$, in the case where it has a single minimum (as shown in Fig.~5, upper left panel):
\begin{align}
|\Psi_N\rangle = \frac{1}{\sqrt{N!}}\left(\cos(\theta/2)\beta_{\Phi+\pi/N_s}^{\dagger}+\sin(\theta/2)\beta_{\Phi-\pi/N_s}^{\dagger}\right)^N|0\rangle.
\end{align}
One readily obtains that the total energy is minimized by the choice $\theta=\pi$  if $1-6u_{\Phi+\pi/N_s}^2v_{\Phi+\pi/N_s}^2<0$, while one has  $\theta=\pi/2$ if $1-6u_{\Phi+\pi/N_s}^2v_{\Phi+\pi/N_s}^2>0$. However, by using the results of Appendix A for the amplitudes $u_k$ and $v_k$, one readily finds that in the Meissner phase $1-6u_{\Phi+\pi/N_s}^2v_{\Phi+\pi/N_s}^2$ is always negative, and we conclude that lowest-energy solution is of biased-ladder type.

\begin{figure}[h!]
\centering
\includegraphics[scale=1]{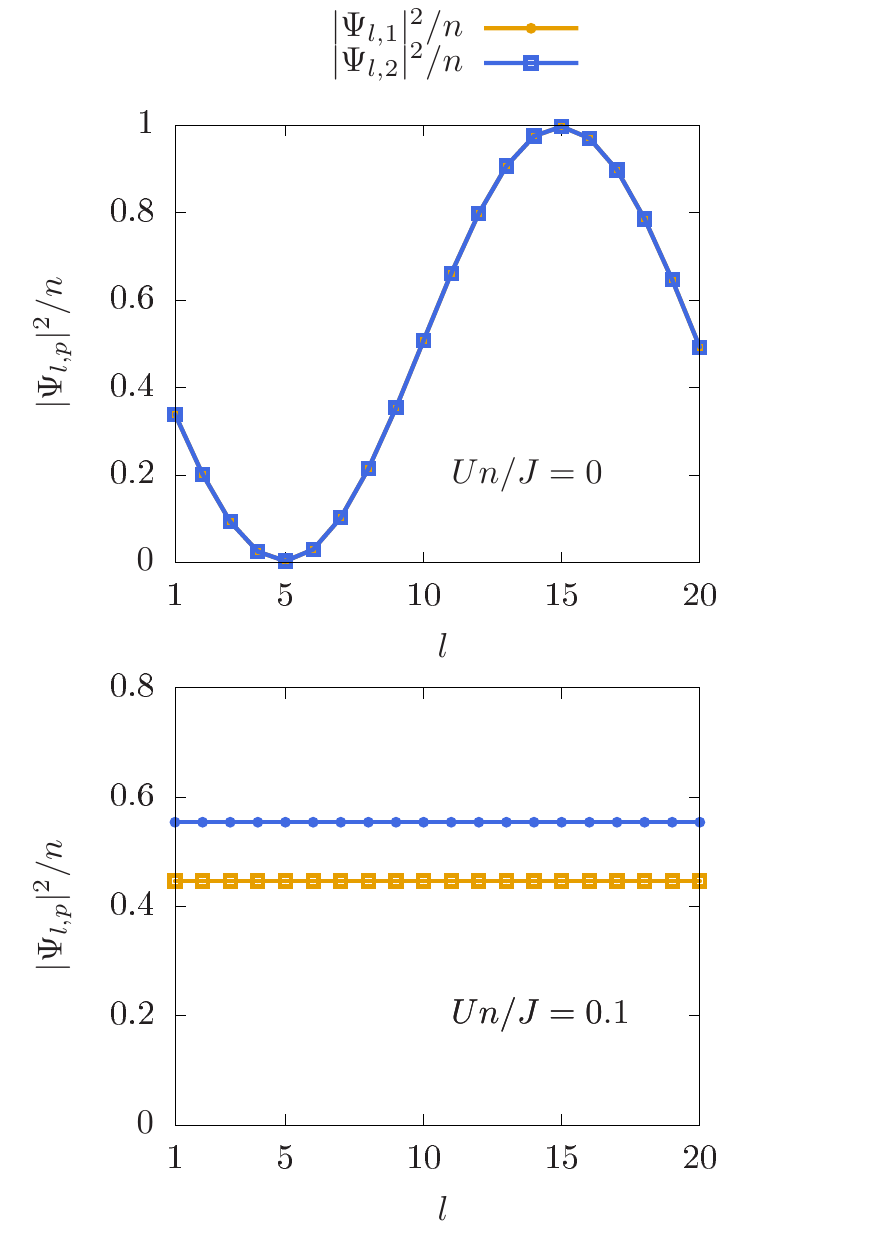}
\caption{(Color online) Density profile for a double ring lattice of interacting bosons  with total flux $\Phi=\pi/N_s$, in the absence of interactions, single vortex in the Meissner phase (upper panel)  and for weak repulsive interactions biased-ladder (BL-M) phase (lower panel). The other parameters are  $N_s=20$,$K/J=2$,$\phi=\pi/2$.}
\label{fig15}
\end{figure}

We have verified this prediction by the numerical solution of the DNLSE, and we confirm that no vortex is found at finite interactions and the density profile is of biased-ladder type, as illustrated in  Fig.~\ref{fig15} and in the phase diagram (Fig.~\ref{Nldif}, lower panel). By performing calculations at varying system size, we find that the imbalance among the two rings decreases with increasing $N_s$.

It is interesting to notice that this is different from the case of the biased ladder phase BL-V obtained for values of flux corresponding to even multiples of $\pi/N_s$. In this case, the particle imbalance does not depend on $N_s$ and the phase is also found in the thermodynamic limit.

%Indeed this additional flux $\Phi=\pi/N_s$ or change of frame doesn't make the system invariant due to the lattice configuration and periodic-boundary conditions, and so adds imbalance in the system. INTERESTIGN BUT WOULD NEED FURTHER CHECKS, I PREFER TO KEEP THIS FOR US

\subsection{Persistent currents for interacting bosons on the double ring lattice }
\begin{figure}[h!]
\includegraphics[scale=1]{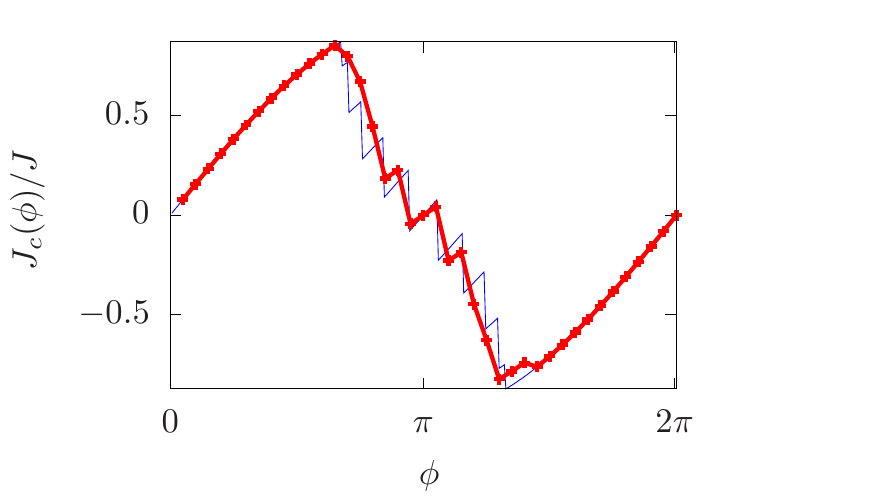}
\caption{(Color online) Chiral currents in units of $J$ as a function of the relative flux $\phi$ (dimensionless)  for noninteracting bosons (blue,  thin solid line) and weakly interacting ones $Un/J=0.1$ (red, thick solid line) for $N_s=20$ and $K/J=3$.}
\label{currents-int}
\end{figure}

The numerical solution of the DNLSE allows also to obtain the persistent currents in the presence of weakly repulsive interactions. Figure \ref{currents-int} shows the dependence on persistent currents amplitude on relative flux $\phi$ for the interacting double ring lattice. As compared to the noninteracting case, notable differences occur at increasing $\phi$ when the phase boundary is crossed: due to the presence of the intermediate biased-ladder phase,  the jumps in the persistent current are suppressed as they are associated to the creation of vortices. For the parameter choice used in Fig.~\ref{currents-int} one can then identify both the transition from Meissner to biased ladder and from the latter to the vortex phase.
Persistent currents thus provide a powerful tool to explore the phases of the double ring lattice.

\section{Spiral interferograms}

\begin{figure}[t]
\includegraphics[scale=1]{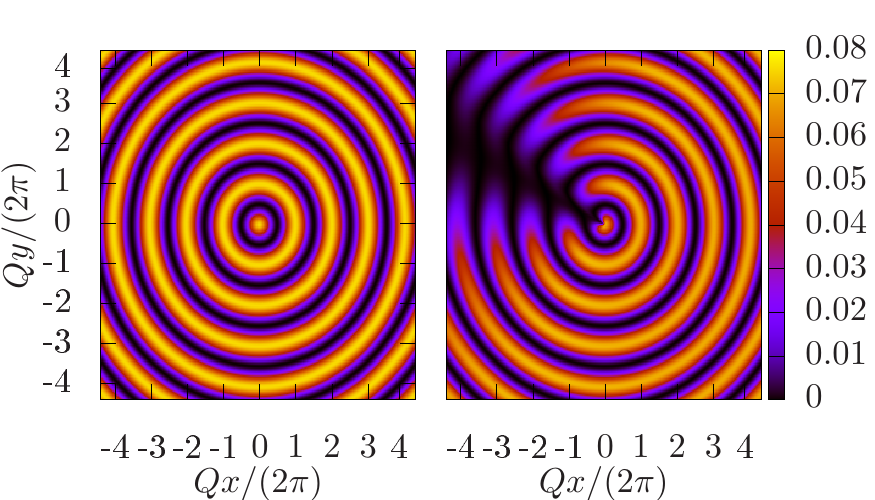}
\includegraphics[scale=1]{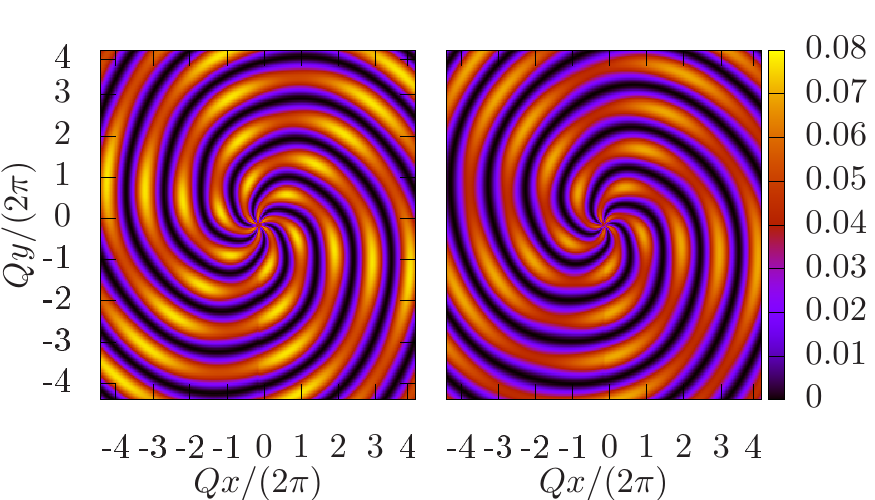}
\caption{(Color online) Spiral interferogram in the Meissner phase (upper panels) with $K/J=1.5$, $\phi=\pi/2$  and $Un/J=0.3$ ,   $\Phi=0$ (upper left panel),  $Un/J=0$, $\Phi=\frac{\pi}{N_s}$ (upper right panel), and in the vortex phase (lower panels) taking $K/J=0.1$,$Un/J=0.3$, $\phi=\pi/3$, with $\Phi=0$ (lower left panel), and  $\Phi=\frac{\pi}{N_s}$ (lower right panel). In all panels  $N_s=35$.
}
\label{OddEven}
\end{figure}
It has been shown~\cite{PhaseInter,PhaseInter2,1367-2630-18-7-075003}  that it is possible to reconstruct the phase pattern of ring trapped Bose-Einstein condensate by studying its interference pattern with a reference disk-shaped condensate placed at the center of the ring. Using a similar principle, we show here that the interference pattern of two concentric rings allows to characterize the vortices in the bosonic double ring lattice.

Assuming that  the distance between neighbouring sites on each ring is  larger than the difference of the radii of the two rings, the main contribution to the interference process is due to radially  overlapping condensates belonging to the same site index in each ring (ie with the same angular coordinate). 
In this case, the wave function after after a time $t_{TOF}$ from  releasing the double ring trap  is given by (see Appendix D for details):
\begin{eqnarray}
\Psi_p(r,\theta_l) \approx \tilde{\Psi}_0(k_{s,p})e^{i\hbar\frac{k_{s,p}^2}{2m}t_{TOF}}e^{i\phi_{l,p}}\sqrt{n_{l,p}}
\label{eq:psip}
\end{eqnarray}
where $\phi_{l,p}$ and $n_{l,p}$  are respectively the phase and the number of particles of a condensate on the ring $p$ at site $l$, and $k_{s,p} = \frac{(R_p-r)(-1)^pm}{\hbar t_{TOF}}$ is related to the velocity at which each wave function evolve after releasing the trap. The interference pattern intensity is given by $I(r,\theta)= 2$ Re$[\Psi^*_1(r,\theta)\Psi_2(r,\theta)]$. By recalling that in density-phase representation $\sqrt{n_{l,1}n_{l,2}}e^{i(\phi_{l,1}-\phi_{l,2})}= \langle a_{l,2}^{\dagger}a_{l,1}\rangle$, we obtain the following intensity distribution in the polar plane $(r,\theta_l)$:
\begin{eqnarray}
I(r,\theta_l) & = &\langle a_{l,1}^{\dagger}a_{l,1}\rangle+\langle a_{l,2}^{\dagger}a_{l,2}\rangle+2 {\rm Re}\left[e^{i\Delta_R}e^{iQr}\langle a_{l,1}^{\dagger}a_{l,2}\rangle\right]\nonumber\\
\label{eq:spirals}
\end{eqnarray}
with $Q=\frac{m(R_1-R_2)}{\hbar t_{TOF}}$ and $\Delta_R = \frac{(R_2^2-R_1^2)m}{\hbar t_{TOF}}$.

In order to analyze typical interference profiles in the various phases, we start from the noninteracting regime. In this case, using the results of Appendix A, in the Meissner phase one readily obtains 
\begin{align}
I(r,\theta_l) \propto \frac{1}{N_s}\cos(Qr+\Delta_R) + n_{\theta_l}
\label{eq:interf}
\end{align}
where $n_{\theta_l}=\langle a_{l,1}^{\dagger}a_{l,1} + a_{l,2}^{\dagger}a_{l,2}\rangle$. This corresponds to an interference pattern made of concentric rings, as illustrated in the first panel of Fig.\ref{OddEven}. 

In the case of a single vortex in the Meissner phase, (second panel of Fig.\ref{OddEven}) the interference pattern displays  a line of dislocations, which are due to the phase slip and  vanishing of the density in correspondence of the vortex core.

In the vortex phase, Eq.(\ref{eq:spirals}) yields 
\begin{eqnarray}
I(r,\theta_l)&\propto& \frac{1}{N_s}[2u_{k_1} v_{k_1}\cos(Qr+\Delta_R)\nonumber\\&+&v_{k_1}^2\cos(\theta_l (k_2-k_1)-\Delta_R - Qr)\nonumber\\&+&u_{k_1}^2\cos(\theta_l (k_2-k_1)+\Delta_R+Qr)]\nonumber\\&+ n_{\theta_l}.
\end{eqnarray}
In this case, the interference pattern  is composed of a term which is constant along  $\theta$, that gives rise to concentric rings and two spirals patterns with uniform intensity each of them corresponding to one of the two ring , one going clockwise and the other counter-clockwise. The superposition of the three contributions  yields a modulated spiral pattern, shown in Fig.\ref{OddEven}. This method, which is specific for the ring geometry,  is a very powerful characterization of the vortex phase, as the number of branches in the pattern yields the number of vortices in the system. This allows in particular to evidence the possibility of having even or odd number of vortices, depending on the value of the total flux. As a final remark we notice that the interference pattern is dependent on the choice of gauge, other choices will lead to different spiral interferogram pictures.

\section{Conclusions and outlook}
In this work, we have studied the ground-state properties of weakly interacting bosons on a double ring lattice, subjected to two gauge fields.
Depending on the ratio between inter-ring and intra-ring tunnel energies, as well on the relative flux, the bosons are found to be in the Meissner or vortex phases, previously identified for the linear ladder geometry.
As specific of the ring geometry, for the non interacting gas, we have found a parity effect on the number of vortices in the system, which originates from the commensurability of total flux with respect to allowed momentum states on the rings. Also, for special values of total flux $\Phi$, due to finite size effects, we have found that the ground state may host  a single vortex even  in the Meissner phase. The analysis of persistent currents shows that at varying relative flux it is possible to identify both the Meissner and vortex phase. In the latter, due to finite-size of the double ring lattice, it is possible to monitor the appearence of pairs of vortices at increasing $\phi$.

We have then considered the effect of weakly repulsive interactions, as described within a mean-field approach.
 We have identified the biased ladder phase and shown that the Meissner phase becomes imbalanced at odd value of the total flux $\Phi$ due to mesoscopic effects. 
Even in the presence of interactions, the study of persistent currents is a useful tool to characterize the various phases.

Finally, we have proposed the interference patterns among the two rings  as probe of the various phases, specifically adapted to the our ring geometry, yielding in particular spiral images in the presence of vortices.

An analysis beyond mean field suggests that the very small ring lattice at weak filling displays fragmentation \cite{Kolovsky} in a  similar way as what is found for spin-orbit coupled Bose gases \cite{Eiji}. In outlook, it would be interesting to explore the crossover from mean-field to fragmented state at decreasing the lattice filling and size.

\acknowledgements
We thank  Luigi Amico, Roberta Citro, Romain Dubessy, Rosario Fazio, Fabrice Gerbier, Erich Mueller, Maxim Olshanii, Paolo Pedri, and  Hélène Perrin for fruitful discussions. We acknowledge funding from the ANR SuperRing project (ANR-15-CE30-0012-02).

\bibliographystyle{prsty}
\bibliography{BibTexNotes}
\newpage
\appendix

\section{Diagonalisation of the non interacting Hamiltonian}

In order to diagonalize the Hamiltonian $H_0$ in Eq.(\ref{eq1}) we introduce the Fourier transform of the field operator according to  $a_{l,p} = \frac{1}{\sqrt{N_s}}\sum_k a_{k,p}e^{-ik_pl}$. Periodic boundary conditions on each ring $ a_{l,p} = a_{l+N_s,p} $ lead to quantized values for the  wavevectors $k= \frac{2\pi}{N_s}j$, where  $j\in \left[0, N_s-1\right]$ is an integer number. 
The Hamiltonian in Fourier space then reads
\begin{eqnarray}
\hat{H_0} = \sum_k 
\left(\begin{matrix}
a^{\dagger}_{k,1} & a^{\dagger}_{k,2}
\end{matrix}
\right)
H(k)
\left(
\begin{matrix}
a_{k,1}\\
a_{k,2}
\end{matrix}
\right),
\end{eqnarray}
where $H(k)$ is given by
\begin{eqnarray}
\!\!H(k)\!=\!
\begin{pmatrix}
\!\!-2J\cos(k\!-\!\Phi -\! \phi/2) & -K \\
-K & \!\!-2J\cos(k\!-\!\Phi +\! \phi/2)
\end{pmatrix}.
\label{hofk}
\end{eqnarray}
We diagonalize it using the unitary transformation
\begin{eqnarray}
\begin{pmatrix}
a_{k,1}\\
a_{k,2}
\end{pmatrix}
=
\begin{pmatrix}
v_k & u_k\\
-u_k & v_k
\end{pmatrix}
\begin{pmatrix}
\alpha_k\\
\beta_k
\end{pmatrix},
\end{eqnarray}
where $u_k$ and $ v_k$ are given by
\begin{eqnarray}
v_k = \sqrt{\frac{1}{2}(1+\frac{\sin(\phi/2)\sin(k-\!\Phi)}{\sqrt{(K/2J)^2+\!\sin^2(\phi/2)\sin^2(k-\Phi))}})}\\
u_k = \sqrt{\frac{1}{2}(1-\frac{\sin(\phi/2)\sin(k-\!\Phi)}{\sqrt{(K/2J)^2+\!\sin^2(\phi/2)\sin^2(k-\Phi))}})}.
\end{eqnarray}
The final form for the Hamiltonian reads
\begin{eqnarray}
\hat{H_0} = \sum_k \alpha_k^{\dagger}\alpha_kE_+(k) + \beta_k^{\dagger}\beta_k E_-(k),
\end{eqnarray}
with 
\begin{eqnarray}
E_{\pm} &=& -2J\cos(\phi/2)\cos(k-\Phi) \\
&\pm& \sqrt{K^2+(2J)^2\sin^2(\phi/2)\sin^2(k-\Phi)}.
\end{eqnarray}

\begin{figure}
\centering
\includegraphics[scale=1]{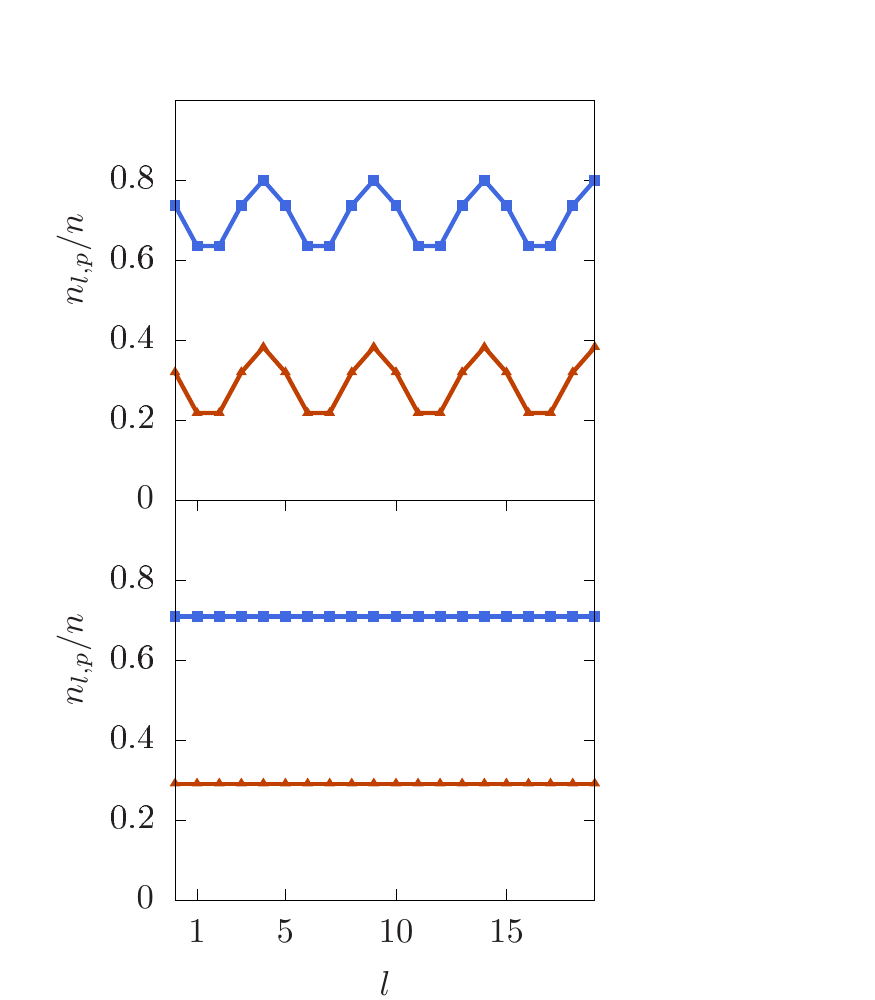}
\caption{Density on the double ring for $K/J=0.95$,$N_s=20$,$\phi=\pi/2$ and upper panel $\theta=0.01$, down panel $\theta=0$.}
\label{Fig:13}
\end{figure}

In the non-interacting regime, in the parameter region where the energy spectrum has a double minimum,   the ground state has the form  $|\psi\rangle = \frac{1}{\sqrt{N!}}\left(\cos(\theta/2)\beta_{k_1}^{\dagger}+\sin(\theta/2)\beta^{\dagger}_{k_2}\right)^N |0\rangle$. This state is fully degenerate in the occupancy of the minima, ie  it provides the same ground-state energy for any choice of $\theta$.  As discussed in section IV, this degeneracy is broken at the level of mean-field by the interactions. In section III we chose to consider, in the non-interacting regime, only the case $\theta = \pi/2$ which leads to the same occupancy of the minima and hence the same density profiles on the two rings. Different choices for $\theta$ will  induce different density profiles, eg an imbalanced vortex  for $\theta \in$ $]0,\pi[$ and $\theta \neq \pi/2$, and a biased-ladder for $\theta=0$ and $\theta=\pi$ (see Fig.~\ref{Fig:13}).
\section{Numerical method for the solution of the DNLSE}

We provide here the details for the numerical solution of the DNLSE (\ref{dnlse}), obtained by iterative steps of the type
\begin{eqnarray}
&&\begin{pmatrix}
\Psi_1(t+\Delta t)\\
\Psi_2(t+\Delta t)
\end{pmatrix}
= U(\Delta t) 
\begin{pmatrix}
\Psi_1(t)\\
\Psi_2(t)
\end{pmatrix}
\end{eqnarray}
where  $U(\Delta t)=\exp(-iH\Delta t)$  is the time-evolution operator and we have introduced the vector notation $\Psi_{p} = \{\Psi_{1,p},...,\Psi_{N_s,p}\}$
Using the Campbell Hausdorff formula we approximate it to order $(\Delta t)^2$ by
\begin{eqnarray}
U(t,t+\Delta t) =
e^{-iH_{0}\Delta t}e^{-iH_{int}\Delta t} + O(\Delta t)^2
\end{eqnarray}
The interacting Hamiltonian being diagonal in position space and the kinetic one in $k$-space, we use the split-step Fourier algorithm ~\cite{Split}. Furthermore,  to obtain the ground-state wave function we perform an evolution in imaginary times. Hence the evolution of our wave function can be recast as follows:
\begin{widetext}
\begin{eqnarray}
&&\begin{pmatrix}
\Psi_1(t+\Delta t)\\
\Psi_2(t+\Delta t)
\end{pmatrix}
=\mathcal{F}^{-1}\left[M\begin{pmatrix}
e^{-E_+\Delta t }\mathbb{I}_{N_s} & 0_{N_s\times N_s}\\
0_{N_s\times N_s} & e^{- E_-\Delta t}\mathbb{I}_{N_s}
\end{pmatrix}M^{-1} \mathcal{F}\left[\begin{pmatrix}
e^{-U|\Psi_{1}(t)|^2 \Delta t}\Psi_1(t) \\
e^{-U|\Psi_{2}(t)|^2 \Delta t}\Psi_2(t)
\end{pmatrix}\right]\right],
\end{eqnarray}
\end{widetext}
where  $|\Psi_{p}(t)|^2 = \{|\Psi_{l,p}|^2, .... |\Psi_{N_s,p}|^2\}$, $E_{\pm} = \{E_{\pm}(2\pi/N_s),...,E_{\pm}(2\pi j/N_s ),...,E_{\pm}(2\pi) \}$, and
$M$ is the unitary matrix which diagonalizes the noninteracting Hamiltonian $H(k)$ (\ref{hofk}) according to
\begin{eqnarray}
M=\begin{pmatrix}
v_k & u_k\\
-u_k & v_k
\end{pmatrix}
\end{eqnarray}
 and $\mathcal{F}$ indicates the Fourier transform.

\section{Interference patterns of expanding rings}

We derive here the expression for the intensity of the interference pattern of expanding rings given in Eq.(\ref{eq:interf}).

We consider first the expansion dynamics of a single condensate initially subjected to a tightly confining potential. We follow  the time evolution of the condensate wavefunction following a sudden turn-off of the confinement at time $t=0$. We will also assume that, due to a sudden decrease of the condensate density, interactions can be neglected during the dynamics, they indeed affect the dynamics of the condensate only in the initial stages of the expansion \cite{CastinDum}. If  $|\Psi(0)\rangle$ is the initial state of the system,  its time evolution following the trap opening is given by 
\begin{eqnarray}
|\Psi(t)\rangle &=& e^{-i H t/\hbar}|\Psi(0)\rangle \nonumber\\
&\simeq & e^{-i H_{kin} t/\hbar} \sum_{\mathbf{k}}|\mathbf{k}\rangle\langle \mathbf{k}|\Psi(0)\rangle\nonumber\\&=&\sum_{\mathbf{k}}\tilde{\Psi}_0(\mathbf{k})e^{-i\frac{\hbar \mathbf{k}^2}{2m}t}|\mathbf{k}\rangle
\end{eqnarray}
where $H_{kin}=\hat p^2/2m$ is the kinetic part of the Hamiltonian.
This readily yields
\begin{eqnarray}
\Psi(\mathbf{x},t)= \int d^2\mathbf{k}\tilde{\Psi}_0(\mathbf{k})e^{-i\frac{\hbar \mathbf{k}^2}{2m}t}e^{i\mathbf{k}\cdot \mathbf{x}}
\end{eqnarray}
Using the saddle-point method to approximate the above integral, in the long-time limit we obtain
\begin{align}
\Psi(\mathbf{x},t) \approx \sqrt{\frac{2\pi m}{t\hbar}}\tilde{\Psi}_0(\bar k_x,\bar k_y)e^{i\frac{m}{2t\hbar}(x^2+y^2)},
\label{eq:tof}
\end{align}
where $\bar k_x=x m/(\hbar t) $, $\bar k_y= y m/(\hbar t)$, thus corresponding to the ballistic regime of the expansion -- the condensate expands at constant velocity, reaching a point in space fixed by its initial momentum in the trap.

In the specific case where the initial confining potential is a double ring lattice, where  $V(\mathbf{x})=\sum_l \frac{1}{2}m\omega^2|\mathbf{x}-\mathbf{x}_l|^2$ and $\mathbf{x}_l$ indicate the minima of the double ring lattice in a two-dimensional plane, we study the expansion and intereference of the condensates released from each ring lattice. Assuming a deep lattice for each ring, and weak inter-ring tunneling, one may consider each lattice site $l$ as occupied by a condensate with phase  $\phi_{l,p}$ and density $n_{l,p}$ weakly coupled to the condensates on the adjacent sites.

After releasing both ring lattices, as well  turning off the artificial gauge fields, using Eq.(\ref{eq:tof}) above, the condensate wavefunctions  will overlap and give rise to an interference pattern at position $r$. If the confinement is very tight in the radial direction or the inter-ring distance is larger than the distance among adjacent sites, the first interference fringes are obtained by the superposition of the condensates wavefunctions radially expanding, ie belonging to the same site index $l$. This is estimated assuming that each condensate  has travelled a distance $(R_p-r)$ at constant velocity $k_{s,p} = (-1)^p (R_p-r)m/\hbar t_{TOF}$, where $p=1,2$ labels each ring, thus acquiring a dynamical phase $\hbar t_{TOF} k_{s,p}^2/2m$, which  adds to the initial phase $\phi_{l,p}$. Taking into account the normalization of each condensate, one readily obtains Eq.(\ref{eq:psip}).  As discussed in \cite{1367-2630-18-7-075003,Luigi-GPE} these interference fringes, and in particular the spirals founds in the vortex phase, occur for typical times $t_{TOF}$  of the order of $\tau_K=m (R_2-R_1) \sigma_r/\hbar $ with $\sigma_r$ the size of the initial condensate in each well. This time is large enough to ensure ballistic expansion (ie on times larger than $\tau_b=m \sigma_r^2/\hbar$), but shorter than the time where neighbouring condensates would contribute to the interference pattern (ie $\tau_J=2\pi m R_p \sigma_r/\hbar N_s$) and would wash out the spirals. At extremely long times,  the time-of-flight images will correspond to the momentum distribution of the initial double ring lattice.

\end{document}